\documentclass[conference]{IEEEtran}

\usepackage{cite}
\usepackage{amsmath,amssymb,amsfonts}
\usepackage{algorithmic}
\usepackage{graphicx}
\usepackage{textcomp}
\usepackage{xcolor}
\def\BibTeX{{\rm B\kern-.05em{\sc i\kern-.025em b}\kern-.08em
    T\kern-.1667em\lower.7ex\hbox{E}\kern-.125emX}}

\usepackage{pgfplots}
\usetikzlibrary{positioning,fit,shapes,calc}
\usepackage{subcaption}
\PassOptionsToPackage{hyphens}{url}\usepackage{hyperref}
\usepackage{multirow}
\usepackage{booktabs}
\usepackage{dblfloatfix}

\definecolor{lirio_main}{HTML}{F06923}
\definecolor{lirio_blue}{HTML}{0A4150}
\definecolor{lirio_red}{HTML}{B9561D}
\definecolor{lirio_orange}{HTML}{F59105}
\definecolor{lirio_gray}{HTML}{C9D4D7}
\definecolor{lirio_lightblue}{HTML}{4A707A}
\definecolor{lirio_yellow}{HTML}{FEC010}

\begin{document}

\title{Mamba for Scalable and Efficient Personalized Recommendations}

\author{\IEEEauthorblockN{Andrew Starnes}
\IEEEauthorblockA{Lirio AI Research \\
Lirio, Inc.\\
Knoxville, TN, USA \\
astarnes@lirio.com}
\and
\IEEEauthorblockN{Clayton Webster}
\IEEEauthorblockA{Lirio AI Research \\
Lirio, Inc.\\
Knoxville, TN, USA \\
cwebster@lirio.com}
}

%\author{\IEEEauthorblockN{Anonymous}
%\IEEEauthorblockA{Anonymous \\
%Anonymous\\
%Anonymous\\
%Anonymous}
%}

\maketitle

\begin{abstract}
In this effort, we propose using the Mamba for handling tabular data in personalized recommendation systems.
We present the \textit{FT-Mamba} (Feature Tokenizer\,$+$\,Mamba), a novel hybrid model that replaces Transformer layers with Mamba layers within the FT-Transformer architecture, for handling tabular data in personalized recommendation systems. The \textit{Mamba model} offers an efficient alternative to Transformers, reducing computational complexity from quadratic to linear by enhancing the capabilities of State Space Models (SSMs). FT-Mamba is designed to improve the scalability and efficiency of recommendation systems while maintaining performance.
We evaluate FT-Mamba in comparison to a traditional Transformer-based model within a Two-Tower architecture on three datasets: Spotify music recommendation, H\&M fashion recommendation, and vaccine messaging recommendation. Each model is trained on 160,000 user-action pairs, and performance is measured using precision (P), recall (R), Mean Reciprocal Rank (MRR), and Hit Ratio (HR) at several truncation values. Our results demonstrate that FT-Mamba outperforms the Transformer-based model in terms of computational efficiency while maintaining or exceeding performance across key recommendation metrics.
By leveraging Mamba layers, FT-Mamba provides a scalable and effective solution for large-scale personalized recommendation systems, showcasing the potential of the Mamba architecture to enhance both efficiency and accuracy.
\end{abstract}


%=====================================================================%
\section{Introduction}
%=====================================================================%

Personalized recommendation systems have revolutionized industries by delivering tailored content and product suggestions based on user preferences and behaviors. However, efficiently processing large-scale data poses significant challenges. Transformers, which have gained popularity for sequential data tasks such as natural language processing (NLP) and computer vision, excel at capturing long-range dependencies between inputs. Yet, they come with a notable drawback—their quadratic complexity with respect to sequence length, which limits scalability for larger datasets \cite{vaswani2017attention,choromanski2020rethinking}.

The \textit{Mamba model} was introduced to address these scalability concerns by reducing the computational complexity to that of State Space Models (SSMs) such as the Structured State Space Sequence Model (S4) while maintaining the performance benefits of Transformers \cite{gu2021efficiently}. The Mamba architecture processes sequential data similarly to Transformers but with several innovations that allow it to achieve linear complexity $\mathcal{O}(L)$ for a sequence length of $L$, as opposed to the quadratic complexity $\mathcal{O}(L^2)$ of standard Transformers \cite{gu2022state, tay2022long}. The architecture comprises two linear projection layers, a convolutional layer, a State Space Model (SSM), and a nonlinear activation function. A key feature of Mamba is its selective SSM, which dynamically adjusts parameters based on input, thus overcoming the SSM's limitation of being unable to forget previous inputs \cite{gu2021efficiently, el-nouby2021training}.

While this dynamic adjustment improves model performance, it introduces a trade-off, slightly reducing the computational efficiency of the underlying linear recurrent neural network (RNN). To mitigate this, Mamba employs a scanning technique that maintains linear time complexity \cite{el-nouby2021training}. In our implementation, we include an optional normalization layer after the final linear projection, which ensures stable embeddings and prevents divergence in experiments \cite{gu2022state}.

Our research focuses on leveraging the Mamba model for \textit{personalized recommendations} based on tabular user data, such as demographic information. Unlike traditional systems, which typically use sequential data such as purchase histories, we handle user data in tabular form. One common approach for handling tabular data in neural networks is the \textit{FT-Transformer}, which uses a feature tokenizer (FT) to convert tabular data into sequences \cite{zhang2022tabtransformer}. However, FT-Transformers face scalability challenges, particularly when dealing with datasets containing a large number of features, a limitation largely attributed to the quadratic complexity of Transformers \cite{vashishth2022scaling}.

To address this, we propose the \textit{FT-Mamba}, a hybrid model that replaces the Transformer layers in FT-Transformers with Mamba layers, thus improving scalability while retaining the ability to process tokenized tabular data. The feature tokenizer in FT-Mamba converts both categorical and numerical tabular data into sequences, which are passed through Mamba layers for efficient sequential processing \cite{zhang2022tabtransformer}. We slightly modify the FT-Transformer by appending the $[CLS]$ token at the end of the token sequence rather than at the beginning, aligning with the Mamba architecture where the last token aggregates information from all preceding tokens \cite{zhang2022tabtransformer, vashishth2022scaling}.

To assess the effectiveness of the FT-Mamba, we conduct experiments using three datasets: Spotify music recommendation, H\&M fashion recommendation, and vaccine messaging recommendation datasets. The models are trained and evaluated, and a range of performance metrics, including precision (P), recall (R), mean reciprocal rank (MRR), and hit ratio (HR), are computed. We compare the FT-Mamba against a traditional Transformer-based model to determine the advantages of Mamba’s architecture in real-world recommendation tasks.

Additionally, we incorporate a \textit{Two-Tower Model} architecture, a common framework for recommendation systems. In this model, one tower encodes user information, and the other encodes content information, with both towers using either FT-Transformer or FT-Mamba layers \cite{yi2019sampling, covington2016deep}. The model’s predictions are based on the inner product of the user and content embeddings, providing a robust and efficient solution for personalized recommendations.  Our results demonstrate that the FT-Mamba consistently outperforms the Transformer-based model in terms of computational efficiency while achieving comparable or superior performance in key recommendation metrics \cite{covington2016deep}.

%=====================================================================%
\section{Methodology}
%=====================================================================%

%---------------------------------------------------------------------%
\subsection{Background on Mamba}
%---------------------------------------------------------------------%

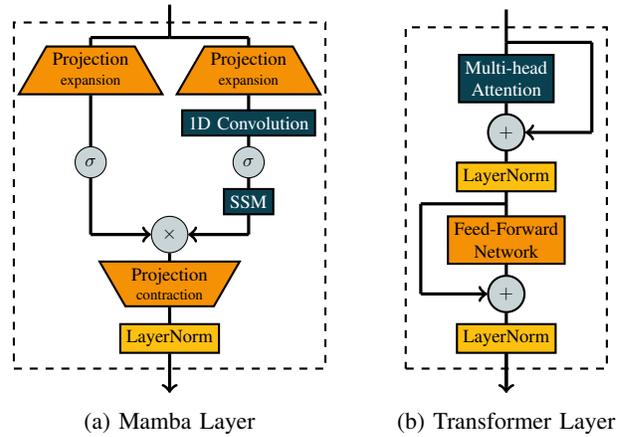
\begin{figure}
    \centering
    \begin{subfigure}[t]{0.24\textwidth}
    \centering
    \pgfdeclarelayer{background layer}
    \pgfdeclarelayer{foreground layer}
    \pgfsetlayers{background layer,main,foreground layer}
    \begin{tikzpicture}[scale=0.7,transform shape]

        \node (C2) at (0,0) {};
        \node[above=0.5cm of C2] (C1) {};
        \node[below=0.5cm of C2, anchor=center, align=center] (C3) {};
        \node[below=0.75cm of C3] (C4) {};
        \node[below=0.5cm of C4] (C5) {};
        \node[below=0.5cm of C5] (C6) {};
        \node[below=0.5cm of C6, draw, circle, anchor=center, align=center, fill = lirio_gray] (C7) {$\times$};
        \node[below=0.5cm of C7.center, draw, thick, trapezium, trapezium left angle=120, trapezium right angle=120, align = center, fill=lirio_orange] (C8) {\shortstack{Projection\\{\footnotesize contraction}}};
        \node[below=0.75cm of C8.center, draw, thick, rectangle, fill = lirio_yellow] (C9) {LayerNorm};
        \node[below=0.5cm of C9] (C10) {};
        
        \node[right=1.5cm of C2.center, anchor = center, align = center] (R2) {};
        \node[right=1.5cm of C3.center,draw, thick, trapezium, trapezium left angle=60, trapezium right angle=60, anchor = center, align = center, fill=lirio_orange] (R3) {\shortstack{Projection\\{\footnotesize expansion}}};
        \node[right=1.5cm of C4.center, draw, thick, rectangle, anchor=center, align=center, fill = lirio_blue] (R4) {\color{white}1D Convolution};
        \node[right=1.5cm of C5.center, draw, circle, anchor=center, align=center, fill = lirio_gray] (R5) {$\sigma$};
        \node[right=1.5cm of C6.center, draw, thick, rectangle, anchor=center, align=center, fill=lirio_blue] (R6) {\color{white} SSM};
        \node[right=1.5cm of C7.center, anchor=center] (R7) {};
        
        \node[left=1.5cm of C2.center, anchor = center, align = center] (L2) {};
        \node[left=1.5cm of C3.center, draw, thick, trapezium, trapezium left angle=60, trapezium right angle=60, anchor=center, align=center, fill=lirio_orange] (L3) {\shortstack{Projection\\{\footnotesize expansion}}};
        \node[left=1.5cm of C5.center, draw, circle, anchor=center, align=center, fill = lirio_gray] (L5) {$\sigma$};
        \node[left=1.5cm of C7.center, anchor=center] (L7) {};

        \draw[very thick] (C1) -- (C2.center);
        \draw[very thick] (C2.center) -- (R2.center);
        \draw[very thick] (C2.center) -- (L2.center);
        \draw[very thick] (L2.center) -- (L3);
        \draw[very thick] (L3) -- (L5);
        \draw[very thick] (L5) -- (L7.center);
        \draw[very thick, ->] (L7.center) -- (C7);
        \draw[very thick] (R2.center) -- (R3);
        \draw[very thick] (R3) -- (R4);
        \draw[very thick] (R4) -- (R5);
        \draw[very thick] (R5) -- (R6);
        \draw[very thick] (R6) -- (R7.center);
        \draw[very thick, ->] (R7.center) -- (C7);
        \draw[very thick] (C7) -- (C8);
        \draw[very thick] (C8) -- (C9);
        \draw[very thick,->] (C9) -- (C10.south);

        \node[left=0.8cm of L2.south west] (B1) {};
        \node[right=0.8cm of R2.south east] (B3) {};
        \begin{pgfonlayer}{background layer}
            \node[draw, thick, inner sep = 2mm, fit=(B1) (C9) (B3), dashed]  {};
        \end{pgfonlayer}
        
    \end{tikzpicture}
    \caption{Mamba Layer}
    \label{fig:mamba}
    \end{subfigure}
    \begin{subfigure}[t]{0.24\textwidth}
    \centering
    \pgfdeclarelayer{background layer}
    \pgfdeclarelayer{foreground layer}
    \pgfsetlayers{background layer,main,foreground layer}
    \begin{tikzpicture}[scale=0.7,transform shape]

        \node (C1) at (0,0) {};
        \node[below=0.5cm of C1] (C2) {};
        \node[below=0.1cm of C2, align = center, draw, thick, rectangle, fill = lirio_blue] (C3) {\color{white}Multi-head\\\color{white}Attention};
        \node[below=0.2cm of C3, align = center, draw, thick, circle, fill = lirio_gray] (C4) {$+$};
        \node[below=0.2cm of C4, align = center, draw, thick, rectangle, fill = lirio_yellow] (C5) {LayerNorm};
        \node[below=0.1cm of C5] (C6) {};
        \node[below=0.1cm of C6, align = center, draw, thick, rectangle, fill = lirio_orange] (C7) {Feed-Forward\\Network};
        \node[below=0.2cm of C7, align = center, draw, thick, circle, fill = lirio_gray] (C8) {$+$};
        \node[below=0.2cm of C8, align = center, draw, thick, rectangle, fill = lirio_yellow] (C9) {LayerNorm};
        \node[below=0.5cm of C9] (C10) {};

        \node[right=1.5cm of C2.center] (R2) {};
        \node[right=1.5cm of C4.center] (R4) {};
        \node[right=1.5cm of C6.center] (R6) {};
        \node[right=1.5cm of C8.center] (R8) {};

        \node[left=1.5cm of C2.center] (L2) {};
        \node[left=1.5cm of C5.center] (L5) {};
        \node[left=1.5cm of C6.center] (L6) {};
        \node[left=1.5cm of C8.center] (L8) {};

        % \draw[very thick] (C1) -- (C2.center);
        % \draw[very thick] (C2.center) -- (C3.center);
        % \draw[very thick] (C3.center) -- (C4);
        \draw[very thick] (C1) -- (C2.center);
        \draw[very thick] (C2.center) -- (C3);
        \draw[very thick] (C3) -- (C4);
        \draw[very thick] (C4) -- (C5);
        \draw[very thick] (C5) -- (C6.center);
        \draw[very thick] (C6.center) -- (C7);
        \draw[very thick] (C7) -- (C8);
        \draw[very thick] (C8) -- (C9);
        \draw[very thick] (C9) -- (C10.center);

        \draw[very thick] (C2.center) -- (R2.center);
        \draw[very thick] (R2.center) -- (R4.center);
        \draw[very thick, ->] (R4.center) -- (C4);
        % \draw[very thick] (C6.center) -- (R6.center);
        % \draw[very thick] (R6.center) -- (R8.center);
        % \draw[very thick, ->] (R8.center) -- (C8);
        \draw[very thick] (C6.center) -- (L6.center);
        \draw[very thick] (L6.center) -- (L8.center);
        \draw[very thick, ->] (L8.center) -- (C8);
        \draw[very thick, ->] (C9) -- (C10.south);

        \begin{pgfonlayer}{background layer}
            \node[draw, thick, inner sep = 2mm, fit=(R2.center) (C9) (L8.center), dashed]  {};
        \end{pgfonlayer}
        
    \end{tikzpicture}
    \caption{Transformer Layer}
    \label{fig:transformer}
    \end{subfigure}
    \caption{Embedding layers}
    \label{fig:embedding_layers}
\end{figure}

The Mamba model \cite{gu2023mamba,dao2024transformers} was introduced to reduce the computational complexity of Transformers while enhancing the performance of State Space Models (SSM), such as the Structured State Space Sequence Model (S4) \cite{gu2021efficiently}.
Similar to Transformers and SSMs, the Mamba processes sequential data, such as tokenized sentences.
The Mamba architecture includes two linear projection layers, a convolutional layer, an SSM, and a nonlinear activation function (see Figure~\ref{fig:mamba}).
The SSM component of the Mamba essentially functions as a linear recurrent neural network (RNN), where $h_t=\overline{A}h_{t-1}+\overline{B}x_t$, with $h$ representing the hidden state and $x$ the input. However, $\overline{A}$ and $\overline{B}$ are not directly optimized.
A key limitation of SSMs is their inability to learn to forget inputs. To address this, Mamba incorporates a selective SSM, where the parameters are dynamically adjusted based on the inputs.
However, this introduces a trade-off, reducing the computational efficiency of the linear RNN. To mitigate this, a scanning technique is applied, which maintains the computation in linear time.
Specifically, for a sequence length of $L$, the computational complexity of both Mamba and SSM is linear ($\mathcal{O}(L)$), whereas the standard Transformer exhibits quadratic complexity ($\mathcal{O}(L^2)$).
Our specific implementation of the Mamba is shown in Figure~\ref{fig:mamba}.
We include an optional normalization layer. However, instead of placing it before the final linear projection, as in the original Mamba architecture, we place it afterward to maintain consistency with the original implementation\footnote{\url{https://github.com/state-spaces/mamba/blob/main/mamba_ssm/modules/mamba_simple.py}}. Omitting it caused the embeddings to grow too large in our experiments.

%---------------------------------------------------------------------%
\subsection{Tokenizer and FT-Mamba}
%---------------------------------------------------------------------%

Our goal is to create personalized recommendations using user information.
Specifically, the user data in our case is tabular (e.g., demographic information) rather than sequential (e.g., purchase history).
The FT-Transformer, introduced in \cite{gorishniy2021revisiting}, utilizes a feature tokenizer (FT) combined with Transformer layers.
The feature tokenizer converts tabular data into sequence-like data that can be processed by the Transformer.
A notable drawback of the FT-Transformer is its limited scalability when handling datasets with a large number of features.
Although this limitation is not solely due to the Transformer component (as discussed in the following footnote), replacing the Transformer with a model that scales more efficiently would be beneficial.
As previously mentioned, the Mamba was proposed as a more scalable alternative to the Transformer.
Therefore, we replace the Transformer layers (Figure~\ref{fig:transformer}) with Mamba layers (Figure~\ref{fig:mamba}).
This model, which combines the feature tokenizer with Mamba layers, is referred to as the FT-Mamba. Below, we detail the feature tokenizer based on \cite{gorishniy2021revisiting}.

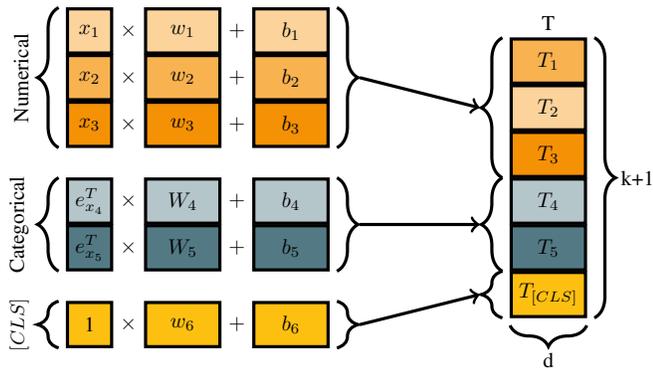
\begin{figure}
    \centering
    \begin{tikzpicture}[scale=0.8,transform shape]

        \node[rectangle, draw, very thick, fill = lirio_orange!40!white] (x1){\parbox[c][5mm][c]{5mm}{\centering $x_1$}};
        \node[below=0cm of x1, rectangle, draw, very thick, fill = lirio_orange!70!white] (x2){\parbox[c][5mm][c]{5mm}{\centering $x_2$}};
        \node[below=0cm of x2, rectangle, draw, very thick, fill = lirio_orange] (x3){\parbox[c][5mm][c]{5mm}{\centering $x_3$}};
        \node[below=0.5cm of x3, rectangle, draw, very thick, fill = lirio_blue!30!white] (x4){\parbox[c][5mm][c]{5mm}{\centering $e_{x_4}^{T}$}};
        \node[below=0cm of x4, rectangle, draw, very thick, fill = lirio_blue!70!white] (x5){\parbox[c][5mm][c]{5mm}{\centering $e_{x_5}^{T}$}};
        \node[below=0.5cm of x5, rectangle, draw, very thick, fill=lirio_yellow] (x6){\parbox[c][5mm][c]{5mm}{\centering 1}};

        \node[right=0cm of x1] (m1) {$\times$};
        \node[right=0cm of x2] (m2) {$\times$};
        \node[right=0cm of x3] (m3) {$\times$};
        \node[right=0cm of x4] (m4) {$\times$};
        \node[right=0cm of x5] (m5) {$\times$};
        \node[right=0cm of x6] (m6) {$\times$};

        \node[right=0cm of m1, rectangle, draw, very thick, fill = lirio_orange!40!white] (w1) {\parbox[c][5mm][c]{10mm}{\centering $w_1$}};
        \node[right=0cm of m2, rectangle, draw, very thick, fill = lirio_orange!70!white] (w2) {\parbox[c][5mm][c]{10mm}{\centering $w_2$}};
        \node[right=0cm of m3, rectangle, draw, very thick, fill = lirio_orange] (w3) {\parbox[c][5mm][c]{10mm}{\centering $w_3$}};
        \node[right=0cm of m4, rectangle, draw, very thick, fill = lirio_blue!30!white] (w4) {\parbox[c][5mm][c]{10mm}{\centering $W_4$}};
        \node[right=0cm of m5, rectangle, draw, very thick, fill = lirio_blue!70!white] (w5) {\parbox[c][5mm][c]{10mm}{\centering $W_5$}};
        \node[right=0cm of m6, rectangle, draw, very thick, fill=lirio_yellow] (w6) {\parbox[c][5mm][c]{10mm}{\centering $w_6$}};

        \node[right=0cm of w1] (p1) {$+$};
        \node[right=0cm of w2] (p2) {$+$};
        \node[right=0cm of w3] (p3) {$+$};
        \node[right=0cm of w4] (p4) {$+$};
        \node[right=0cm of w5] (p5) {$+$};
        \node[right=0cm of w6] (p6) {$+$};

        \node[right=0cm of p1, rectangle, draw, very thick, fill = lirio_orange!40!white] (b1) {\parbox[c][5mm][c]{10mm}{\centering $b_1$}};
        \node[right=0cm of p2, rectangle, draw, very thick, fill = lirio_orange!70!white] (b2) {\parbox[c][5mm][c]{10mm}{\centering $b_2$}};
        \node[right=0cm of p3, rectangle, draw, very thick, fill = lirio_orange] (b3) {\parbox[c][5mm][c]{10mm}{\centering $b_3$}};
        \node[right=0cm of p4, rectangle, draw, very thick, fill = lirio_blue!30!white] (b4) {\parbox[c][5mm][c]{10mm}{\centering $b_4$}};
        \node[right=0cm of p5, rectangle, draw, very thick, fill = lirio_blue!70!white] (b5) {\parbox[c][5mm][c]{10mm}{\centering $b_5$}};
        \node[right=0cm of p6, rectangle, draw, very thick, fill=lirio_yellow] (b6) {\parbox[c][5mm][c]{10mm}{\centering $b_6$}};

        \draw [very thick,decorate,decoration={brace,amplitude = 8pt, raise = 0.1cm}] (b1.north east) -- (b3.south east);
        \draw [very thick,decorate,decoration={brace,amplitude = 8pt, raise = 0.1cm}] (b4.north east) -- (b5.south east);
        \draw [very thick,decorate,decoration={brace,amplitude = 8pt, raise = 0.1cm}] (b6.north east) -- (b6.south east);
        
        \draw [very thick,decorate,decoration={brace,mirror,amplitude = 8pt, raise = 0.1cm}] (x1.north west) -- (x3.south west);
        \draw [very thick,decorate,decoration={brace,mirror,amplitude = 8pt, raise = 0.1cm}] (x4.north west) -- (x5.south west);
        \draw [very thick,decorate,decoration={brace,mirror,amplitude = 8pt, raise = 0.1cm}] (x6.north west) -- (x6.south west);

        \node[right=3cm of b4.north east, draw, very thick, fill = lirio_orange, rectangle, anchor = south west] (y3) {\parbox[c][5mm][c]{10mm}{\centering $T_3$}};
        \node[above=0cm of y3, draw, very thick, fill = lirio_orange!40!white, rectangle] (y2) {\parbox[c][5mm][c]{10mm}{\centering $T_2$}};
        \node[above=0cm of y2, draw, very thick, fill = lirio_orange!70!white, rectangle] (y1) {\parbox[c][5mm][c]{10mm}{\centering $T_1$}};
        \node[below=0cm of y3, draw, very thick, fill = lirio_blue!30!white, rectangle] (y4) {\parbox[c][5mm][c]{10mm}{\centering $T_4$}};
        \node[below=0cm of y4, draw, very thick, fill = lirio_blue!70!white, rectangle] (y5) {\parbox[c][5mm][c]{10mm}{\centering $T_5$}};
        \node[below=0cm of y5, draw, very thick, fill=lirio_yellow, rectangle] (y6) {\parbox[c][5mm][c]{10mm}{\centering $T_{[CLS]}$}};

        \draw [very thick,decorate,decoration={brace,mirror,amplitude = 8pt, raise = 0.1cm}] (y1.north west) -- (y3.south west);
        \draw [very thick,decorate,decoration={brace,mirror,amplitude = 8pt, raise = 0.1cm}] (y4.north west) -- (y5.south west);
        \draw [very thick,decorate,decoration={brace,mirror,amplitude = 8pt, raise = 0.1cm}] (y6.north west) -- (y6.south west);

        \node[right=0.38cm of b2] (l2) {};
        \node[right=0.38cm of b5.north east] (l4) {};
        \node[right=0.38cm of b6] (l6) {};

        \node[left=0.38cm of y2] (r2) {};
        \node[left=0.38cm of y5.north west] (r4) {};
        \node[left=0.38cm of y6] (r6) {};

        \draw[very thick, ->] (l2.center) -- (r2.center);
        \draw[very thick, ->] (l4.center) -- (r4.center);
        \draw[very thick, ->] (l6.center) -- (r6.center);

        \draw [very thick,decorate,decoration={brace,amplitude = 8pt, raise = 0.1cm}] (y1.north east) -- (y6.south east);
        \node[right = 0.45cm of y3.south east] {k+1};
        \draw [very thick,decorate,decoration={brace,mirror,amplitude = 8pt, raise = 0.1cm}] (y6.south west) -- (y6.south east);
        \node[below = 0.45cm of y6] {d};
        \node[above=0cm of y1] {T};

        \node[left=0.25cm of x2, label = {[anchor=south,rotate=90]left:Numerical}] {};
        \node[left=0.25cm of x4.south west, label = {[anchor=south,rotate=90]left:Categorical}] {};
        \node[left=0.25cm of x6, label = {[anchor=south,rotate=90]left:{\small$[CLS]$}}] {};
        
    \end{tikzpicture}
    \caption{Image derived from \cite{gorishniy2021revisiting} that visually explains the tokenizer that includes $[CLS]$ token.}
    \label{fig:tokenizer}
\end{figure}

The feature tokenizer converts tabular data into a sequence of tokens, handling categorical and numerical values separately (see Figure~\ref{fig:tokenizer} for an overview).
Let $d$ denote the dimension of the token space, $\mathbb{R}^d$.
For numeric input, the real values are linearly scaled into $\mathbb{R}^d$ as follows: $x_j\mapsto x_jw_j + b_j$, where $w_j, b_j \in \mathbb{R}^d$ are trainable parameters.
For categorical input, the categories of each feature are enumerated, and the feature's index is used instead of the actual value.
Specifically, if $x_j$ is categorical and feature $j$ has $K_j$ categories, then $x_j \in {0, 1, \dots, K_j}$.
For each categorical feature, a lookup table is created where each feature value corresponds to a specific row. We denote the lookup table for feature $j$ as $W_j \in \mathbb{R}^{K_j \times d}$.\footnote{This highlights a scaling issue with the feature tokenizer: memory requirements increase with both the number of features and the number of categories per feature.}
The token for $x_j$ is the value in row $x_j$ of the $j$-th lookup table, plus a bias term (i.e., $e_{x_j}^T W_j + b_j$, where $e_{x_j} \in \mathbb{R}^{K_j}$ is 1 at the $x_j$ coordinate and 0 elsewhere).
In summary, the token for $x_j$ is given by
\begin{equation}
    T_j
    =\left\{\begin{array}{ll}
        x_jw_j+b_j&\text{if }x_j\text{numerical}\\
        e_{x_j}^TW_j+b_j&\text{if }x_j\text{categorical}
    \end{array}\right.
    \in\mathbb{R}^d
\end{equation}
where $w_j, b_j \in \mathbb{R}^d$ and $W_j \in \mathbb{R}^{K_j \times d}$ are all trainable parameters.
Since the tokenizer produces a single embedding vector to represent all of $x$ after passing through the embedding layers (either Mamba or Transformer layers), we introduce a final numeric value of 1 and obtain its token (acting as the $[CLS]$ token \cite{kenton2019bert}), denoted by $T_{[CLS]}=w_{k+1}+b_{k+1}$.
Finally, the representation of $x$ in tokens is given by
\begin{equation}
    T=\text{stack}\Big[T_1,...,T_k,T_{[CLS]}\Big].
\end{equation}
Once the tokens pass through the Mamba or Transformer layers, the resulting embedding from the $[CLS]$ token is used to represent $x$.
A minor deviation from the original FT-Transformer is that we append the $[CLS]$ token at the end of the token sequence instead of at the beginning. In the Mamba, the first token’s output contains only information about the first token, while in both the Transformer and Mamba, the last token contains information from all previous tokens.

%---------------------------------------------------------------------%
\subsection{Two-Tower Models}
%---------------------------------------------------------------------%

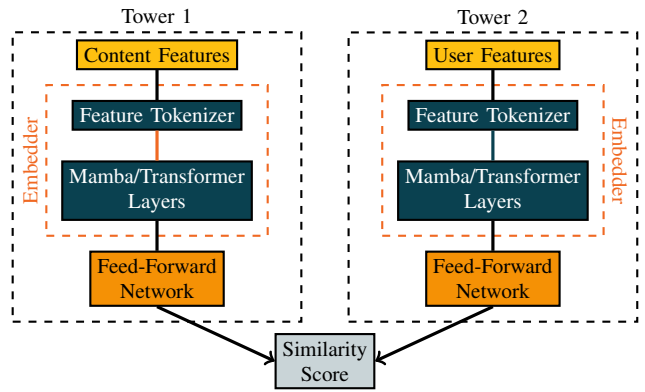
\begin{figure}
    \centering
    \pgfdeclarelayer{background layer}
    \pgfdeclarelayer{foreground layer}
    \pgfsetlayers{background layer,main,foreground layer}
    \begin{tikzpicture}[scale=0.8,transform shape]
        \node (top_center) {};
    
        \node[rectangle, draw, thick, left=1.25cm of top_center, fill = lirio_blue] (cft) {\color{white}Feature Tokenizer};
        \node[rectangle, draw, thick, above=0.5cm of cft, fill = lirio_yellow] (content) {Content Features};
        \node[rectangle, draw, thick, below=0.5cm of cft, align = center, fill = lirio_blue] (cmtl) {\color{white}Mamba/Transformer\\\color{white}Layers};
        \node[rectangle, draw, thick, below=0.5cm of cmtl, align = center, fill = lirio_orange] (cffn) {Feed-Forward\\Network};
        \draw[very thick] (content) -- (cft);
        \draw[very thick, lirio_main] (cft) -- (cmtl);
        \draw[very thick] (cmtl) -- (cffn);
    
        \node[rectangle, draw, thick, right=1.25cm of top_center, fill = lirio_blue] (uft) {\color{white}Feature Tokenizer};
        \node[rectangle, draw, thick, above=0.5cm of uft, fill = lirio_yellow] (user) {User Features};
        \node[rectangle, draw, thick, below=0.5cm of uft, align = center, fill = lirio_blue] (umtl) {\color{white}Mamba/Transformer\\\color{white}Layers};
        \node[rectangle, draw, thick, below=0.5cm of umtl, align = center, fill = lirio_orange] (uffn) {Feed-Forward\\Network};
        \draw[very thick] (user) -- (uft);
        \draw[very thick, lirio_blue] (uft) -- (umtl);
        \draw[very thick] (umtl) -- (uffn);
    
        \node[rectangle, draw, thick, below=3.5cm of top_center, align = center, fill = lirio_gray] (dot) {Similarity\\Score};
        \draw[very thick, ->] (cffn.south) -- (dot.west);
        \draw[very thick, ->] (uffn.south) -- (dot.east);

        \begin{pgfonlayer}{background layer}
            \node[left=1.9cm of content.center] (CL1) {};
            \node[right=1.9cm of content.center] (CR1) {};
            \node[draw, thick, inner sep = 2mm, fit=(CL1) (cffn) (CR1), dashed, label = Tower 1]  {};

            \node[left=1.9cm of user.center] (UL1) {};
            \node[right=1.9cm of user.center] (UR1) {};
            \node[draw, thick, inner sep = 2mm, fit=(UL1) (uffn) (UR1), dashed, label = Tower 2]  {};
            
            \node[draw, thick, inner sep = 2mm, label = {[anchor=south,rotate=270]right:{\color{lirio_main}Embedder}}, fit=(uft) (umtl), dashed, lirio_main]  {};
            \node[draw, thick, inner sep = 2mm, label = {[anchor=south,rotate=90]left:{\color{lirio_main}Embedder}}, fit=(cft) (cmtl), dashed, lirio_main]  {};
        \end{pgfonlayer}
    \end{tikzpicture}
    \caption{Two-Tower Model}
    \label{fig:two_tower_model}
\end{figure}

A Two-Tower Model \cite{huang2013learning,neculoiu2016learning,covington2016deep,yi2019sampling} consists of two independent models (the two towers) that combine to produce the model’s output, and is commonly used in recommendation systems \cite{de2024personalized,yang2020mixed,dixith2024handling,yu2021dual,li2022inttower}.
In our case, one tower encodes user information, while the other encodes content information, making this a more specific use of a dual encoder model \cite{gillick2018end}.
Since both the user and content data are in tabular formats, we use either FT-Transformers or FT-Mamba for the two towers.
The inner product of the user and content embeddings serves as the model's prediction, though this is not the only possible approach for comparing embedding similarity (e.g., cosine similarity or another feed-forward network could be used).
The architectures of the Two-Tower Transformer and the Two-Tower Mamba are illustrated in Figure~\ref{fig:two_tower_model}.

%=====================================================================%
\section{Experiments}
%=====================================================================%

The experiments compare two Two-Tower Models: one using a Transformer as the embedder and the other using a Mamba. The three datasets utilized are the Spotify music recommendation, H\&M fashion recommendation, and vaccine messaging recommendation datasets. Each training dataset contains 160,000 user-action pairs with a target to predict, while the test datasets consist of 100 users. Models are trained using mean squared error (MSE) with a batch size of 32, resulting in one epoch with 5,000 steps.

Each of the feed-forward networks in the Two-Tower Models consists of a hidden layer with 64 nodes and an output layer with 32 nodes, both utilizing ReLU activation. The tokenizer uses tokens of size 192. A hyperparameter search was conducted using the Spotify music recommendation dataset to tune the embedders, with 100 seeds and 1,000 validation points. These tuned hyperparameters were then used in the H\&M and vaccine messaging experiments, with only the learning rates varied, which again were chosen through validation using a few seeds. The Transformer and Mamba towers were configured to use identical hyperparameters between the user and content towers. The Transformer was configured with 2 heads, 0\% head dropout, and 2 layers. The Mamba was set with a convolution width of 16, a linear expansion factor of 2, and 4 layers. Any other hyperparameters used are their default values. This results in the Transformer embedder having 1,186,048 trainable parameters in each tower, compared to the 479,232 trainable parameters in the Mamba embedders. As we will see, despite having only around 40\% of the trainable parameters of the Transformer embedding, the Two-Tower Mamba is often able to outperform the Two-Tower Transformer.

To compute test set metrics, we use 100 different seeds to estimate statistical significance. The metrics computed include precision (P), recall (R), mean reciprocal rank (MRR), and hit ratio (HR), which we evaluate at truncatation values $k=1$, 5, and 10, denoted as P@$k$, R@$k$, MRR@$k$, and HR@$k$, respectively.
Since P@1, MRR@1, and HR@1 are all equal, we only present HR@1.

All experiments were conducted using Google Colab \cite{googlecolab} with a T4 runtime type. The code is available at 
% \url{https://github.com/acstarnes/recommendation}
\url{https://github.com/} (truncated for anonymity), where a link to run the code on Google Colab can be found.

%---------------------------------------------------------------------%
\subsection{Spotify Music Recommendation}
%---------------------------------------------------------------------%

\begin{figure*}[b]
    \centering
    \begin{subfigure}{0.45\linewidth}
        \centering
        \includegraphics[width=0.99\linewidth]{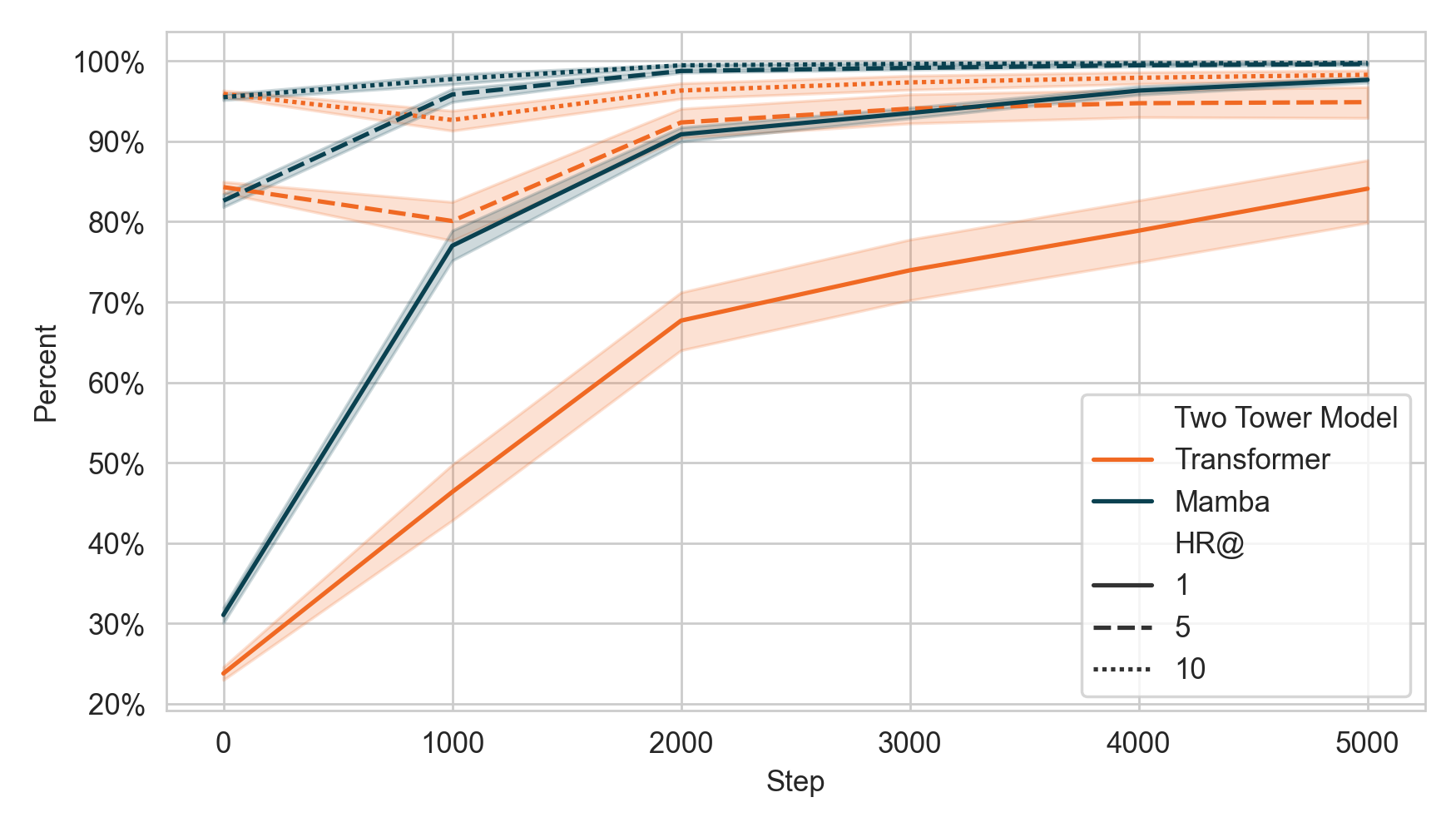}
        \caption{Spotify Music Recommendation}
        \label{fig:spotify_topn}
    \end{subfigure}
    \begin{subfigure}{0.45\linewidth}
        \centering
        \includegraphics[width=0.99\linewidth]{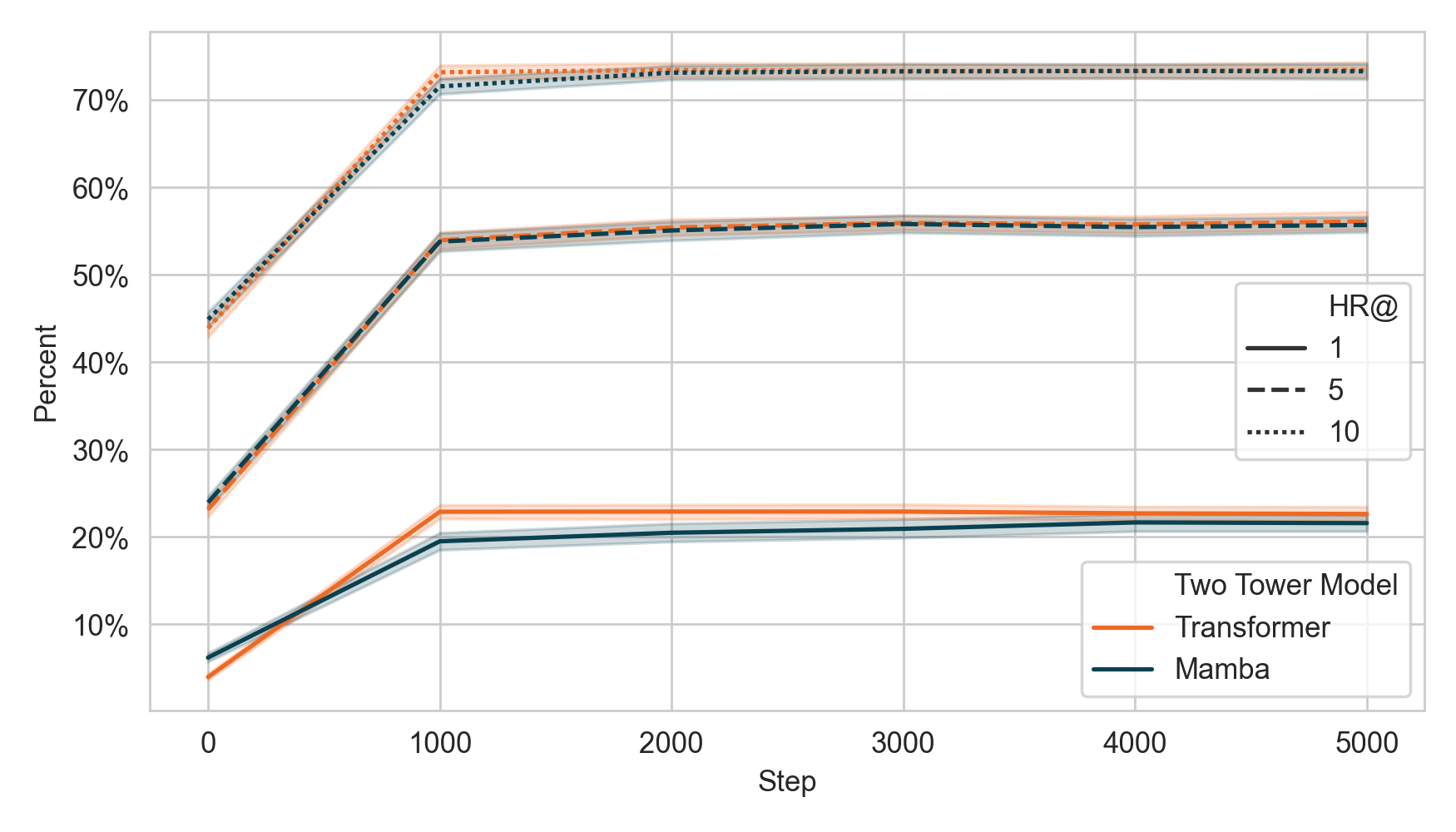}
        \caption{H\&M Fashion Recommendation}
        \label{fig:hm_fashion}
    \end{subfigure}
    \caption{Hit Ratios Through Training}
    \label{fig:enter-label}
\end{figure*}

We use the music recommendation dataset from \cite{dereventsov2022examining}. Users are represented by a 20-dimensional binary vector, with each element indicating a preference for one of 20 different genres. These vectors are treated as numeric features in the tokenizer. The content is represented by the song’s rank on the ``Top 50-Global'' playlist, which is treated as a categorical feature. The task is to predict a value of $-1$, $0$, or $1$, representing negative, neutral, or positive user sentiment towards the song. For analysis, we focus on how well the models recommended songs the user liked, consolidating the neutral and dislike categories.

In this dataset, each of the 20 genres and 50 songs is represented by a 10-dimensional vector, with coordinates corresponding to the following 10 attributes: Acoustic, Dance, Energy, Instruments, Liveness, Loudness, Mode, Speech, Tempo, and Valence. Thus, genres form a matrix $G \in \mathbb{R}^{20 \times 10}$, and songs form a matrix $S \in \mathbb{R}^{50 \times 10}$. A user's preference for a song is calculated as $p(u) = u G S^T$, where $(p(u))_i$ represents the preference for the $i$-th song in the ``Top 50-Global'' playlist. A threshold is then applied to convert these preferences into $-1$, $0$, or $1$.

Initially, this dataset was designed for reinforcement learning, but we adapt it for a supervised learning setting. To generate the training and validation sets, we randomly sample users and songs, then compute the users' sentiments about the selected songs. For the test set, we randomly sample users and evaluate their sentiments toward all songs on the playlist. 
We conducted a hyperparameter search to optimize both the Two-Tower Transformer and Two-Tower Mamba models, and both of the models use a learning rate of $10^{-4}$.

The metrics for this experiment are presented in the top rows of Table~\ref{tab:spotify}, with Hit Ratios (HR@) throughout training displayed in Figure~\ref{fig:spotify_topn}.
Both models perform exceptionally well, even when recommending just one song.
Notably, the top Two-Tower Mamba recommendation was liked by users an impressive 97.7\% of the time, compared to 84.11\% for the Two-Tower Transformer.
After only 2,000 training steps, both models achieved over 90\% HR@5.
However, across all metrics, Two-Tower Mamba consistently outperformed the Two-Tower Transformer.
Particularly, based on the MRR, we expect the Two-Tower Mamba to recommend songs that the user enjoys quicker than the Two-Tower Transformer.

% \begin{figure}[b]
%     \centering
%     \includegraphics[width=0.99\linewidth]{images/spotify_hit_ratio_single.png}
%     \caption{Spotify music Hit Ratios through training}
%     \label{fig:spotify_topn}
% \end{figure}

\begin{figure*}
    \centering
    \begin{subfigure}{\linewidth}
        \centering
        \includegraphics[width=0.99\linewidth]{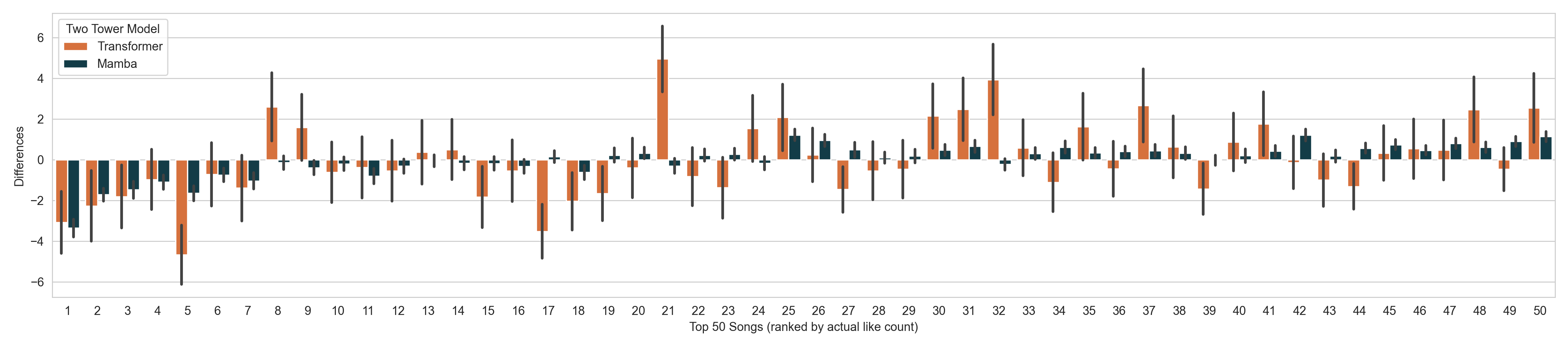}
        \caption{Spotify Music Recommendation}
        \label{fig:spotify_diff}
    \end{subfigure}
    \begin{subfigure}{\linewidth}
        \centering
        \includegraphics[width=0.98\linewidth]{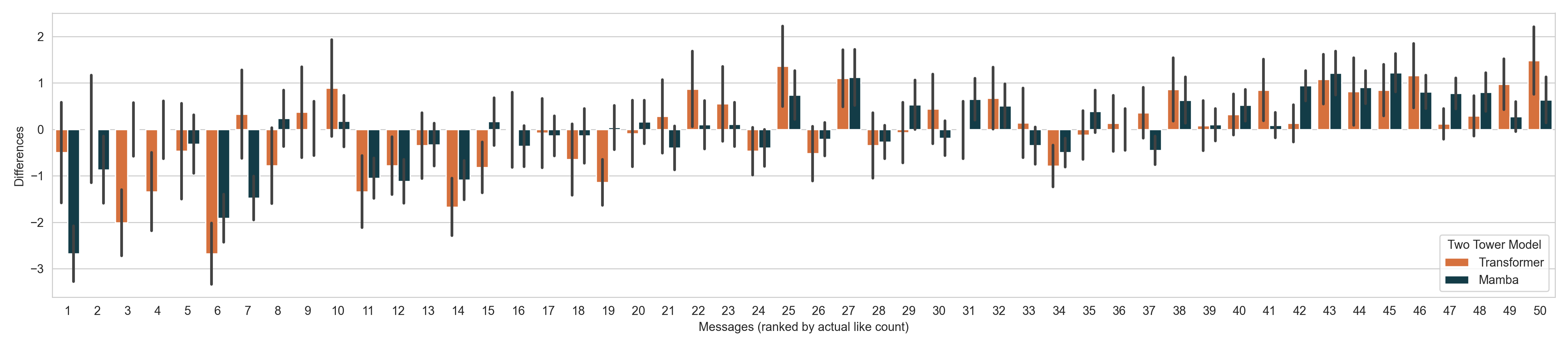}
        \caption{Messaging Recommendation}
        \label{fig:messaging_diff}
    \end{subfigure}
    \caption{Differences between predicted and actual likes, ranked by number of likes per item per seed. Positive values indicate over-recommendations and negative values indicate under-recommendation.}
    \label{fig:diff}
\end{figure*}

Regarding the diversity of song recommendations, Figure~\ref{fig:spotify_diff} highlights the difference between the number of times each song was predicted versus how often users liked those songs.
The number of recommendations for a user is tied to the number of songs that the user liked, which relates to recall, where truncation depends on the individual user.
A value below 0 means the song was liked more often than it was recommended, values above 0 indicate the song was predicted more than it was liked, and values near 0 suggest the recommendation and like counts were similar.
To standardize across seeds, songs were ranked based on the number of user likes rather than playlist order, meaning as rank increases from 1 to 50, the number of likes decreases.

Given the performance metrics already discussed, we anticipate the Two-Tower Transformer's recommendations to deviate further from 0 compared to those of the Two-Tower Mamba.
Nevertheless, a trend emerges in which both models tend to recommend top-rated songs less frequently than they were actually liked.
The Two-Tower Transformer shows a slight tendency to over-recommend certain songs (e.g., ranks 21 and 32).
Importantly, neither model exhibits the issue of only recommending a small subset of songs, a problem previously observed with reinforcement learning approaches on this dataset \cite{dereventsov2022examining}.

% \begin{table}[]
%     \centering
%     \begin{tabular}{|c|c|c|c|c|c|}\hline
%     Model&\multirow{2}{*}{P@10}&\multirow{2}{*}{R@10}&\multicolumn{2}{|c|}{Hit Ratio}&\multirow{2}{*}{MRR}\\
%     Embedder&&&\multicolumn{1}{|c}{HR@1}&\multicolumn{1}{c|}{HR@10}&\\\hline
%     Mamba&\textbf{0.747}&\textbf{0.543}&\textbf{0.977}&\textbf{0.998}&\textbf{0.877}\\
%     Transformer&0.57&0.408&0.841&0.983&0.739\\\hline
%     \end{tabular}
%     \caption{Spotify Accuracy}
%     \label{tab:spotify}
% \end{table}

\begin{table*}
    \centering
    \begin{tabular}{cc|cc|cc|cc|ccc}
\multirow{2}{*}{\parbox{0.07\linewidth}{\centering\vspace{14pt} \textbf{Dataset}}}
&\multicolumn{1}{c}{\multirow{2}{*}{\parbox{0.07\linewidth}{\centering\vspace{5pt} \textbf{Embedder}\\\textbf{Model}}}}
&\multicolumn{2}{c}{\textbf{Precision}}
&\multicolumn{2}{c}{\textbf{Recall}}
&\multicolumn{2}{c}{\textbf{Mean Reciprocal Rank}}
&\multicolumn{3}{c}{\textbf{Hit Ratio}}\\\cmidrule(lr){3-4}\cmidrule(lr){5-6}\cmidrule(lr){7-8}\cmidrule(lr){9-11}
&
&\textbf{P@5}
&\textbf{P@10}
&\textbf{R@5}
&\textbf{R@10}
&\textbf{MRR@5}
&\textbf{MRR@10}
&\textbf{HR@1}
&\textbf{HR@5}
&\textbf{HR@10}\\\specialrule{.1em}{.05em}{.05em}
Spotify&Mamba&0.952&0.906&0.374&0.671&0.985&0.985&0.977&0.996&0.998\\
Spotify&Transformer&0.803&0.759&0.303&0.551&0.882&0.887&0.841&0.949&0.983\\\specialrule{.1em}{.05em}{.05em}
H\&M&Mamba&0.1566&0.1283&0.1207&0.1979&0.3335&0.3569&0.2155&0.5567&0.7326\\
H\&M&Transformer&0.1584&0.1279&0.1223&0.1974&0.3416&0.3646&0.226&0.5609&0.7341\\
    \end{tabular}
    \caption{Spotify Music and H\&M Fashion Performance Metrics}
    \label{tab:spotify}
    \label{tab:hm_fashion}
\end{table*}

% \begin{figure*}
%     \centering
%     \includegraphics[width=0.99\linewidth]{images/spotify_barplot_diff.png}
%     \caption{Spotify Differences}
%     \label{fig:spotify_diff}
% \end{figure*}

%---------------------------------------------------------------------%
\subsection{H\&M Fashion Recommendation}
%---------------------------------------------------------------------%

Our next recommendation environment is based on the H\&M Personalized Fashion dataset \cite{h-and-m-personalized-fashion-recommendations}. Users are represented by one numerical feature (age) and four categorical features: fashion news subscription status, communication activity, club membership status, and frequency of reading fashion news. The content consists of articles of clothing, described by 11 categorical features related to their type and appearance. The original dataset contains over 100,000 items which consumed the memory. Instead, we reduce to the 100 most popular and only use the article name as the sole content categorical feature.

The H\&M Personalized Fashion dataset contains only transaction data, meaning it lacks information on items users never purchased. To address this, we sample one item each user did not purchase and label it as a non-transaction. For transactions, the target label is 1, while for non-transactions, it is 0. Additionally, we limit our test set to users with at least 5 purchases.

The training set consists of 160,000 user-content-response triplets, with roughly half representing transactions and the other half non-transactions. The test set includes 100 users and a list of items each user actually purchased. The objective is to recommend these items to the users. We use a learning rate of $10^{-4}$ for both models.

As shown in Figure~\ref{fig:hm_fashion}, the precision, Mean Reciprocal Rank (MRR), and Hit Ratios during training demonstrate an initial improvement in both models, but minimal or no progress after 1,000 training steps. Additional performance metrics are provided in Table~\ref{tab:hm_fashion}. While the performance of both models leaves room for improvement, the Two-Tower Transformer slightly outperformed the Two-Tower Mamba when focusing on single recommendations (i.e., P@1, MRR@1, and HR@1). Performance could be enhanced by more intelligently selecting non-transactions rather than doing so randomly. Moreover, including additional training data or recommending broader clothing categories instead of specific items may yield better results. The H\&M Personalized Fashion Kaggle leaderboard\footnote{\href{https://www.kaggle.com/competitions/h-and-m-personalized-fashion-recommendations/leaderboard}{https://www.kaggle.com/competitions/h-and-m-personalized-fashion-recommendations/leaderboard}} ranks submissions based on Mean Average Precision @ 12 (MAP@12). The Two-Tower Transformer and Two-Tower Mamba achieved MAP@12 scores of 0.0335 and 0.0342, respectively, which would place these models within the top 20, assuming the results extend to the full dataset.

% \begin{figure}[b]
%     \centering
%     \includegraphics[width=0.99\linewidth]{images/hm_fashion_hit_ratio_single.png}
%     \caption{H\&M Fashion Hit Ratio through training}
%     \label{fig:hm_fashion}
% \end{figure}

% \begin{figure*}
%     \centering
%     \includegraphics[width=0.98\linewidth]{images/hm_fashion_performance.png}
%     \caption{H\&M Personalized Fashion Performance}
%     \label{fig:hm_fashion}
% \end{figure*}

%---------------------------------------------------------------------%
\subsection{Vaccine Messaging Recommendation}
%---------------------------------------------------------------------%

\begin{figure*}[b]
    \centering
    \includegraphics[width=0.98\linewidth]{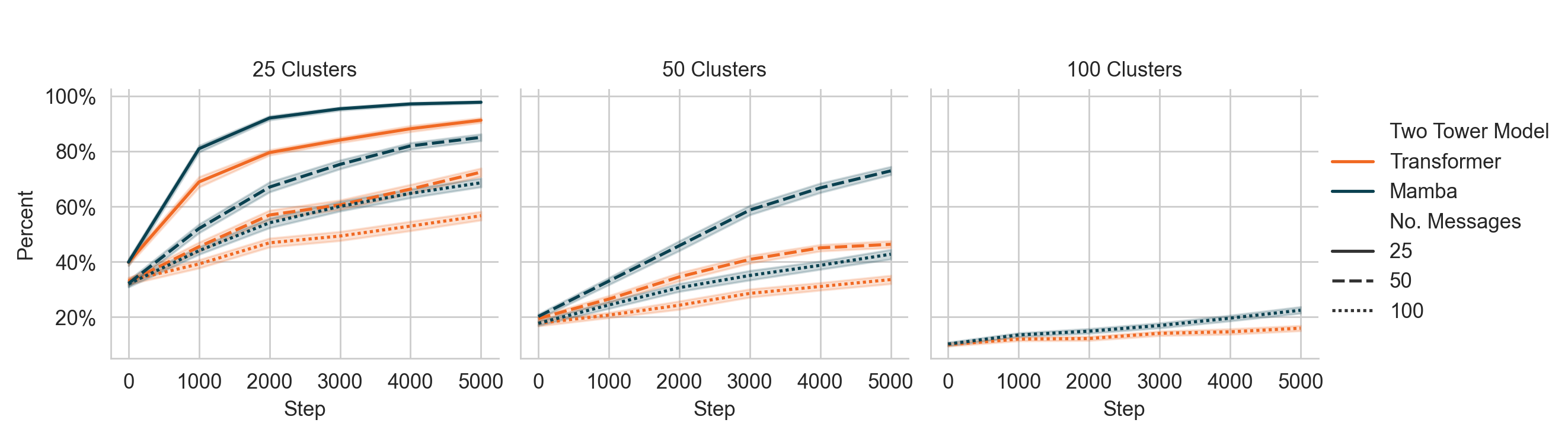}
    \caption{Vaccine Messaging Hit Ratio (HR@10) for varying numbers of clusters and messages}
    \label{fig:messaging_hit_ratio}
\end{figure*}

The final personalization environment is a synthetic setting where user information is available, and the model's task is to select content to encourage vaccination. 
Users are characterized by one numerical feature (age) and seven categorical features: gender, progression, education, insurance, race, language, and religion. These features are randomly sampled.
The content comprises 100 messages generated by ChatGPT \cite{chatgpt} using the prompt: ``You are responsible for creating messages to encourage people to take a COVID-19 vaccine. Please provide 100 unique messages.'' 
These messages are processed through an Instructor-based LLM \cite{su2022one}, producing 768 numerical features for each message.

K-means clustering, with $k=25,50,100$, is applied separately to both user and content features, creating (at most) $k$ distinct clusters for each. 
We also experiment with varying the number of messages, so that we look at $k=25$ with 25, 50, and 100 messages, $k=50$ with 50 and 100 messages, and $k=100$ with 100 messages, where the messages are randomly sampled each seed.
Users who receive content from the same cluster are more likely to take the vaccine (target = 1), while those receiving content from different clusters are less likely (target = 0).
The training dataset includes 160,000 messages from 50,000 users, with random sampling that may result in duplicates. 
The test set consists of 100 users and a list of the content most likely to prompt vaccination.
During validation, we found that representing the index of the messages performed better than using the Instructor embedding numerical features.
This is likely due to the fact that the number of messages is substantially less that the Instructor embeddings (i.e., 100 $<$  768).
We also found a learning rate of $10^{-5}$ worked best on the validations sets.

The performance metrics are presented in Table~\ref{tab:messaging}, and additional details on the Hit Ratio during training are shown in Figure~\ref{fig:messaging_hit_ratio}. Once again, the Two-Tower Mamba outperforms the Two-Tower Transformer across all metrics. Notably, the HR@1 for the Two-Tower Mamba is almost twice that of the Two-Tower Transformer for most cluster-message combinations. In practical terms, this suggests that the Two-Tower Mamba would result in nearly double the number of vaccinations after the first message compared to the Two-Tower Transformer.

As the number of clusters increases, the models' performance declines, and the same trend occurs if the number of messages increases while keeping clusters fixed. This behavior is somewhat expected, as additional clusters or messages introduce more complexity, requiring more training data to capture the user-message relationship accurately. Furthermore, with more clusters, there are fewer positive user-message interactions in each batch, suggesting that increasing the batch size could be beneficial.

Similar to the Spotify music recommendation experiment, Figure~\ref{fig:messaging_diff} shows the difference between predicted and actual likes (i.e., the message had a target of 1), with messages ordered by their actual number of likes to allow comparison across seeds. Although the plot shows only 50 actions and 25 clusters, the observed trends apply to other combinations of actions and clusters. First, we see a tendency to underpredict messages with ranks close to 1 and 50. Second, neither model significantly over- nor under-recommends any particular message. Therefore, if deployed in a production setting, this model would generate a diverse set of recommendations, covering a broad range of user-message pairs.

% \begin{figure*}
%     \centering
%     \includegraphics[width=0.98\linewidth]{images/messaging_barplot_diff_50a_25c.png}
%     \caption{Messaging Difference}
%     \label{fig:messaging_diff}
% \end{figure*}

\begin{table*}[]
    \centering
    \begin{tabular}{ccc|cc|cc|cc|ccc}
\multirow{2}{*}{\parbox{0.08\linewidth}{\centering\vspace{5pt} \textbf{Number of}\\\textbf{Clusters}}}
&\multirow{2}{*}{\parbox{0.08\linewidth}{\centering\vspace{5pt} \textbf{Number of}\\\textbf{Messages}}}
&\multicolumn{1}{c}{\multirow{2}{*}{\parbox{0.07\linewidth}{\centering\vspace{5pt} \textbf{Embedder}\\\textbf{Model}}}}
&\multicolumn{2}{c}{\textbf{Precision}}
&\multicolumn{2}{c}{\textbf{Recall}}
&\multicolumn{2}{c}{\textbf{Mean Reciprocal Rank}}
&\multicolumn{3}{c}{\textbf{Hit Ratio}}\\\cmidrule(lr){4-5}\cmidrule(lr){6-7}\cmidrule(lr){8-9}\cmidrule(lr){10-12}
&
&
&\textbf{P@5}
&\textbf{P@10}
&\textbf{R@5}
&\textbf{R@10}
&\textbf{MRR@5}
&\textbf{MRR@10}
&\textbf{HR@1}
&\textbf{HR@5}
&\textbf{HR@10}\\\specialrule{.1em}{.05em}{.05em}
\multirow{6}{*}{25}&\multirow{2}{*}{25}&Mamba&0.1794&0.0977&0.897&0.977&0.6738&0.6848&0.5351&0.897&0.977\\
                                       &&Transformer&0.1362&0.0912&0.681&0.9119&0.3925&0.4236&0.2355&0.681&0.9119\\\cline{2-12}
                   &\multirow{2}{*}{50}&Mamba&0.2597&0.1718&0.6269&0.8219&0.4823&0.5033&0.3719&0.6939&0.8498\\
                                       &&Transformer&0.1733&0.1366&0.4212&0.6558&0.3167&0.3448&0.2159&0.5129&0.7241\\\cline{2-12}
                   &\multirow{2}{*}{100}&Mamba&0.2592&0.22&0.2939&0.4822&0.3779&0.3985&0.2945&0.5296&0.6856\\
                                       &&Transformer&0.1536&0.14&0.1896&0.3286&0.2519&0.2733&0.1725&0.4042&0.5664\\\specialrule{.1em}{.05em}{.05em}
\multirow{4}{*}{50}&\multirow{2}{*}{50}&Mamba&0.1046&0.0729&0.523&0.7288&0.2986&0.3262&0.1801&0.523&0.7288\\
                                       &&Transformer&0.0508&0.0463&0.254&0.4629&0.121&0.1481&0.0568&0.254&0.4629\\\cline{2-12}
                  &\multirow{2}{*}{100}&Mamba&0.0884&0.076&0.1843&0.3187&0.1705&0.189&0.1106&0.2863&0.4272\\
                                       &&Transformer&0.0498&0.0471&0.1187&0.2275&0.0997&0.1179&0.0512&0.1957&0.3354\\\specialrule{.1em}{.05em}{.05em}
\multirow{2}{*}{100}&\multirow{2}{*}{100}&Mamba&0.0256&0.0224&0.1281&0.2245&0.0638&0.0763&0.0323&0.1281&0.2245\\
                                       &&Transformer&0.0176&0.016&0.0878&0.1595&0.0425&0.0517&0.0204&0.0878&0.1595\\
    \end{tabular}
    \caption{Vaccine Messaging Performance Metrics}
    \label{tab:messaging}
\end{table*}

%=====================================================================%
\section{Related Works}
%=====================================================================%

Two recent models also handle passing tabular data into the Mamba, but neither focuses on recommendations as we do here.
First, MambaTab \cite{ahamed2024mambatab} processes tabular data by encoding the features and then applying a feed-forward neural network to embed the values.
Categorical features are encoded by enumerating the categories and using the index of each category as its feature value.
The categorical features are then combined with the numeric features and passed through a normalization layer, which serves as input to the embedding neural network.
Although this method imposes an arbitrary ordering on the categories, the authors found that it does not negatively affect performance.
Their experiments focus on supervised, self-supervised, and feature-incremental learning tasks.
Second, Mambular \cite{thielmann2024mambular} was developed concurrently with our work.
Similar to FT-Mamba, Mambular combines the FT-Transformer with the Mamba but utilizes periodic encodings (from \cite{gorishniy2022embeddings}, co-authored by some of the original FT-Transformer contributors) for their numeric features.
n both classification and regression tasks, Mambular achieves similar or improved performance compared to the FT-Transformer and consistently outperforms MambaTab.
However, in classification tasks, XGBoost often delivered the best performance.

Since Transformers are commonly used in recommendation systems (see \cite{pohan2022recommender} for a survey), replacing them with the Mamba, similiar to what we have done here, is a natural next step.
While no other papers focus on the specific type of recommendations we explore here, three have used the Mamba for sequential recommendations.
\cite{liu2024mamba4rec} employs the selective SSM for sequential recommendations, demonstrating state-of-the-art performance compared to other systems.
However, they include a feed-forward neural network after what we call the Mamba layer here, making their Mamba layer quite similar to a Transformer layer.
\cite{zhang2024matrrec}, proposed a few months later, combines the Mamba and Transformer for recommendations, giving rise to the MaTrRec moniker.
Specifically, the MaTrRec block consists of a Mamba block followed by a Transformer block.
Their experiments found that the Mamba has lower recall rates on short sequences.
They hypothesized that the Mamba excels at capturing long-sequence dependencies, while the Transformer is better suited for short sequences.
In the experiments, MaTrRec outperformed other recommendation systems, with Mamba4Rec often being the runner-up.
Finally, the most recent approach is introduced by \cite{liu2024bidirectional}.
They introduce the Partially Flipped Mamba (PF-Mamba), which introduces bidirectional flow into the Mamba by partially flipping the first several terms of the interaction sequence.
The PF-Mamba consists of two Mamba blocks: one standard Mamba block and one that incorporates flipping.
Additionally, to improve modeling of short-term patterns in interaction sequences, they introduce the Feature Extract Gated Recurrent Unit (FE-GRU), which integrates a convolutional layer with a GRU.
These components combine to form their SelectIve Gated MAmba (SIGMA).
In their experiments, SIGMA consistently delivered the best performance, always outperforming Mamba4Rec.
Notably, MaTrRec was not included in the experiments.

%=====================================================================%
\section{Conclusion}
%=====================================================================%

In this work we replaced Transformer layers with Mamba layers, and demonstrating how the linear computational complexity of Mamba improves the overall scalability of recommendation models, particularly when dealing with large-scale tabular data.
In addition, we exploited the \textit{FT-Mamba}, a hybrid model that integrates the Mamba architecture with the FT-Transformer to address scalability and efficiency challenges in personalized recommendation systems. 

Our experiments, conducted on three diverse datasets-Spotify music recommendation, H\&M fashion recommendation and vaccine messaging recommendation—show that FT-Mamba consistently outperforms a traditional Transformer-based model in terms of computational efficiency, while maintaining or exceeding performance across several key metrics, including precision, recall, Mean Reciprocal Rank, and Hit Ratio. These results highlight the effectiveness of Mamba's architecture in reducing computational overhead without sacrificing accuracy.

The findings from this study suggest that Mamba offers a promising direction for future developments in scalable machine learning models, particularly for real-time applications requiring personalized recommendations. As recommendation systems continue to evolve with increasingly larger datasets, the FT-Mamba model presents a viable solution for improving both model efficiency and user experience. Future work may focus on exploring additional optimizations to the Mamba architecture and testing its application across broader datasets and real-world recommendation tasks.

%=====================================================================%
% \section*{Acknowledgment}
%=====================================================================%

%=====================================================================%
\bibliographystyle{plain}
\bibliography{main}
%=====================================================================%

\end{document}